\documentclass[journal]{IEEEtran}

\usepackage{cite}
\usepackage{bookmark}
\ifCLASSINFOpdf
   \usepackage[pdftex]{graphicx}
   \graphicspath{{../pdf/}{../jpeg/}}
   \DeclareGraphicsExtensions{.pdf,.jpeg,.png}
\else
   \usepackage[dvips]{graphicx}
   \graphicspath{{../eps/}}
   \DeclareGraphicsExtensions{.eps}
\fi

\usepackage{multirow}
\usepackage{xcolor}
\usepackage{amsmath}
\usepackage{dsfont}
\usepackage{amsfonts}
\usepackage{algorithmic}

\newenvironment{algogo}[1]{
\smallskip
\noindent \hrule\vspace{0.2\baselineskip} \hrule
\begin{small}
\refstepcounter{algo} \center{\bf \textsc{Algorithm \thealgo}}
\\{\center{\bf #1}}
\smallskip
\flushleft
 } {
\end{small}
\smallskip
\hrule\vspace{0.2\baselineskip} \hrule
\smallskip
}

\newcounter{algo}
\renewcommand{\thealgo}{\arabic{algo}}

\ifCLASSOPTIONcaptionsoff
  \usepackage[nomarkers]{endfloat}
 \let\MYoriglatexcaption\caption
 \renewcommand{\caption}[2][\relax]{\MYoriglatexcaption[#2]{#2}}
\fi

\newcommand{\bPhi}{\boldsymbol{\Phi} }

\newcommand{\blue}{\textcolor{blue} }

\begin{document}
\title{EM-based approach to 3D reconstruction from single-waveform multispectral Lidar data}

\author{Quentin~Legros,
         Sylvain~Meignen,
        Stephen~McLaughlin, Yoann~Altmann
\thanks{This work was supported by the Royal Academy of Engineering under the Research Fellowship scheme RF201617/16/31 and by the Engineering and Physical Sciences Research Council (EPSRC) (grant EP/S000631/1) and the MOD University Defence Research Collaboration (UDRC) in Signal Processing. and the IDEX of University Grenoble Alpes.}
\thanks{Quentin Legros, Yoann Altmann and Stephen McLaughlin are with the School of Engineering anf Physical Sciences, Heriot Watt University, Edinburgh, United Kingdom). Sylvain Meignen is with the Laboratoire Jean Kuntzmann, Grenoble, France.}
}


%
%

\maketitle

\begin{abstract}
In this paper, we present a novel Bayesian approach for estimating spectral and range profiles from single-photon Lidar waveforms associated with single surfaces in the photon-limited regime. In contrast to classical multispectral Lidar signals, we consider a single Lidar waveform per pixel, whereby a single detector is used to acquire information simultaneously at multiple wavelengths.
A new observation model based on a mixture of distributions is developed. It relates the unknown parameters of interest to the observed waveforms containing information from multiple wavelengths. 
Adopting a Bayesian approach, several prior models are investigated and a stochastic Expectation-Maximization algorithm is proposed to estimate the spectral and depth profiles. The reconstruction performance and computational complexity of our approach are assessed, for different prior models, through a series of experiments using synthetic and real data under different observation scenarios. The results obtained demonstrate a significant speed-up without significant degradation of the reconstruction performance when compared to existing methods in the photon-starved regime. 
\end{abstract}


\IEEEpeerreviewmaketitle

\section{Introduction}








\IEEEPARstart{D}{epth} measurement from Light detection and ranging (Lidar) systems has received an increasing interest over the last decades, as it allows reconstruction of 3D scenes at high resolution\cite{BoschThierry2001Lrac, Lidar_system_2018}. Single band Lidar (SBL), i.e., using a single illumination wavelength, allows the extraction of spatial structures from 3D scenes, using a pulsed illumination and analysis of the time-of-arrival (ToA) of detected photons, for each spatial location or pixel. More precisely, time correlated single-photon counting (TCSPC) correlates the photon ToAs with the time of emission of the pulses to obtain the time-of-flights (ToFs) of the recorded photons that are gathered to form a histogram of photon detection events. Neglecting light scattering in the medium, the detected photons originally emitted by the pulsed source are temporally clustered and form a peak in the histogram, whose amplitude informs on the reflectivity of the surface. Conversely, the additional detection events caused by ambient illumination and detector dark counts are classically uniformly distributed across the histogram bins. While allowing the reconstruction of high-resolution 3D scenes, SBL only provides one reflectivity value (corresponding to the wavelength of the laser source) per surface and thus does not provide spectral information about the scene of interest. When such information is also required, data fusion strategies can be adopted, as in \cite{Fusion_1,Fusion_2} for instance.

Multispectral Lidar (MSL) overcomes data registration and fusion problems, as spectral and spatial features are collected using a single imaging system \cite{wallace_buller_Lidar}. This emerging modality has already led to promising results for underwater imaging \cite{UnderwaterImaging}, object detection in urban area \cite{UrbanArea} and forest canopy monitoring \cite{Forest-Canopy}.
Using laser sources at different wavelengths, histograms associated with different spectral bands are usually recorded for each pixel of the scenes, using a single device. 
The acquisition of MSL signals can be performed either sequentially, i.e., one wavelength at a time, or in parallel. Sequential acquisition requires an overall longer acquisition time which depends on the number of wavelengths sensed. While parallel acquisition allows faster acquisition, it requires a more complex and potentially more expensive imaging system \cite{MSL_simultaneous_acquisition,MSL_mul_sensor_corresp,MSL_mul_sensor}, e.g., using fibers optic \cite{Optic_fib} or optical volume gratings \cite{opt_vol_gratings} to separate the spectral components.

\begin{figure}[htb]
 \begin{center}
\resizebox{88mm}{!}{
\includegraphics{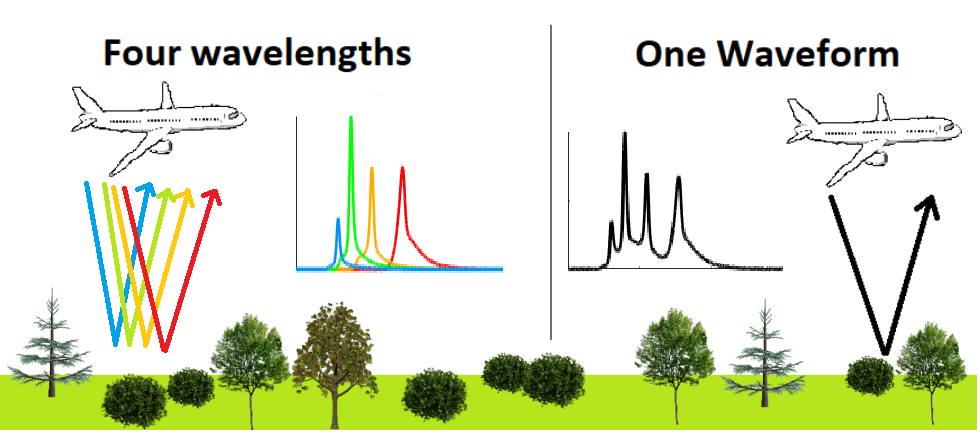}
}
 \end{center}
 \caption{Examples of acquisition of four MSL waveforms (left), using one wavelength per waveform, and SW-MSL (right) capturing simultaneously information from four wavelengths. In this example, the vegetated scene is mostly green and the Lidar return associated with this wavelength range (second peak from the left) is larger than the other returns.
 \label{fig:One_Waveform_acquisition}}
\end{figure}

To reduce the acquisition time, limit the complexity of the imaging system and the volume of data to be stored and processed, a single-waveform MSL (SW-MSL) approach has recently been developed \cite{Comp_realData}, whereby a single histogram containing spectral information from multiple wavelengths is recorded per pixel via wavelength-time coding (see Fig \ref{fig:One_Waveform_acquisition}). Using a single waveform per pixel, the acquisition time is of the same order as when using SBL in the photon-starved regime. 
As discussed in \cite{Comp_realData}, the distribution of the photons ToA is shifted in time with a wavelength-dependent temporal delay. Although these delays are fixed, the amplitude of the different peaks varies depending on the spectral signatures of the scene, which changes the waveform shape.

The study of SW-MSL data reduces to estimating a 3D profile and to quantifying the proportion of detected photons associated with each wavelength. In a similar fashion to the analysis of MSL data, the joint estimation of the spectral and range profiles is a challenging problem, mainly due to the multimodal nature of the observation model. 
To overcome this difficulty, Bayesian approaches have been proposed in the MSL case, e.g.,  \cite{Tobin_SPIRAL,Journal_Spectral_unmixing_MSL_Yoann,MuSaPoP,halimi2019robust} 
for their ability to explore multimodal distributions and to quantify the uncertainty associated with the estimates (e.g., range profiles).

To analyse SW-MSL data, the Bayesian method proposed in \cite{Comp_realData} relies also on Monte Carlo sampling to estimate the spectral and depth profiles. Although it provides promising reconstruction results, this method suffers from a prohibitive computational time (about 15 hours to reconstruct a single 3D scene of $200 \times 200$ pixels and $4$ wavelengths), which motivates the development of new and more scalable methods based on optimization schemes, as intended here.


In this work, we introduce a new SW-MSL model within a Bayesian framework and an associated inference procedure which is significantly faster than that proposed by Ren et al. \cite{Comp_realData}. The classical observation model based on Poisson noise as in \cite{Multispec_Yoann1,model_Bay1,Journal_Spectral_unmixing_MSL_Yoann,BayesianMCMC_Lidar,Tachella2019}, is replaced by an equivalent model as in \cite{EusipcoYoann} (in the low-flux regime) which is more suited for the proposed inference strategy. More precisely, the distribution of the photon detection events is represented by a mixture model, where one component of the mixture relates to the ambient illumination (background) and where the remaining components are associated with the different illumination wavelengths (signal contributions). Similarly to most 3D reconstruction methods from single-photon data, such as \cite{EusipcoYoann,BayesianMCMC_Lidar,Journal_Spectral_unmixing_MSL_Yoann,Tobin_SPIRAL}, we assume the photons emitted by the imaging system are reflected onto a single target. 
We then assign prior distributions to the unknown model parameters. To overcome convergence issues (multimodal posterior distribution, slow convergence of the MCMC method in \cite{Comp_realData}) induced by the joint estimation of the depth and spectral profile, we estimate the parameters sequentially. More precisely, we use an algorithm based on {\it Expectation-Maximization} (EM) to first estimate the mixture weights in each pixel. These weights are then used to estimate spectral signature in each pixels and the range profile.


The main contributions of the paper are:
\begin{itemize}
    \item A new SW-MSL observation model whose formulation is well suited for inference using EM-based algorithms. 
    \item An EM-based algorithm for analysis of SW-MSL data which extends the model proposed in \cite{EusipcoYoann} for SBL data. 
    \item In contrast to the method proposed in \cite{EusipcoYoann}, we provide an additional step to estimate reflectivity parameters from the estimated ratios of informative photons detected.
    \item A comparison of several prior models for the unknown spectral profile in terms of reconstruction performance and associated computational complexity. This study demonstrates that the proposed strategy yields faster reconstruction than the method in \cite{Comp_realData}.

\end{itemize}


The remainder of this paper is organized as follows. Section \ref{Sec:Model} introduces the new SW-MSL observation model and the different prior distributions associated with the unknown model parameters are discussed in Section \ref{Sec:Priors}. Section \ref{Sec:Strategy} describes the estimation strategy based on EM.
A comparison of the different prior models using synthetic SBL data is conducted in Section \ref{Sec:Results} in order to identify the most interesting options in terms of estimation performance and computational cost. This study is then extended in the remainder of Section \ref{Sec:Results}, which investigates real SW-MSL data analysis and comparison with the method proposed in \cite{Comp_realData}. Conclusions and future work are finally reported in Section \ref{Sec:Conclusion}.


\section{Problem Formulation}
 \label{Sec:Model}

Consider a 3D observation array $\boldsymbol{Z}$ of MSL waveforms $\boldsymbol{z}_{n,l}= [\boldsymbol{Z}]_{n,l,:}=[z_{n,l,1},\ldots,z_{n,l,T}]^{\top}$ of size $N \times L \times T$, where $N$ is the number of pixels or spatial locations, $L$ is the number of spectral bands and $T$ is the length of the ToF histograms. The spectral response of the object in the $n$th pixel is denoted by $\boldsymbol{r}_{n} = [r_{n,1},\dots,r_{n,L}]^{\top} \in \mathds{R}_{+}^{L}$, and $b_{n,l}$ is the positive, constant over time, background level of that pixel, at the $l$th wavelength. The classical MSL observation model relies on Poisson noise \cite{Multispec_Yoann1, Tobin_SPIRAL} such that 
\begin{equation}\label{eq:model_poisson_MSL}
    z_{n,l,t}| \left(r_{n,l},t_{n},b_{n,l} \right) \sim \mathcal{P}\left( r_{n,l}  g_{l}(t - t_{n}) + b_{n,l}\right),
\end{equation}
where $\{g_{l}(.)\}_{l}$ are the instrumental/impulse response functions (IRFs) associated with the $L$ wavelengths.
These IRFs are assumed to be known as they are generally estimated during the system calibration. Moreover, $t_{n}$ is the characteristic ToF of photons emitted by the laser source and reflected by an object at distance $d_{n}$ of the sensor. Note that $d_{n}$ and $t_n$ are linearly related when the light speed is constant in the medium between the imaging device and the scene. 

In a similar manner, the SW-MSL observation model based on Poisson noise can be obtained by summing the elements of $\boldsymbol{Z}$ over the spectral dimension (see also \cite{Comp_realData}), that is, by defining $y_{n,t}=\sum_{l=1}^L z_{n,l,t}$, resulting in an $N \times T$ observation matrix $\boldsymbol{Y}$ with $\left[\boldsymbol{Y}\right]_{n,t}=y_{n,t}$. For convenience, we also define $\boldsymbol{y}_{n} = [y_{n,1},\dots,y_{n,T}]^{\top}$, the waveform associated with the $n$th pixel, such that $\boldsymbol{Y}=[\boldsymbol{y}_{1},\ldots,\boldsymbol{y}_{N}]^{\top}$.
The resulting SW-MSL observation model becomes
\begin{equation}\label{eq:model_poisson}
    y_{n,t}| \left(\boldsymbol{r}_{n},t_{n},b_{n}\right) \sim \mathcal{P}\left( b_{n} + \sum_{l=1}^{L} r_{n,l}  g_{l}(t - t_{n}) \right),
\end{equation}
where $b_{n}$ gathers all the background contributions in the $n$th pixel.
The main challenge with the model in Eq. \eqref{eq:model_poisson} is the joint estimation of $\boldsymbol{t} = \{t_{n}\}_{n}$ and $(\boldsymbol{R}, \boldsymbol{b}) = (\{\boldsymbol{r}_{n}\}_{n}, \{b_{n}\}_{n})$, especially for high background levels that lead to a highly multimodal likelihood in Eq. \eqref{eq:model_poisson}.

Our approach uses a model reformulation similar to that in \cite{EusipcoYoann}, as it allows the use of EM-based algorithms for faster parameter estimation. Instead of using the waveforms $\boldsymbol{y}_{n}$ as observations, it uses lists of photon ToFs.
We denote by $\boldsymbol{s}_{n}=\{s_{n}^{p}\}_{p=1}^{\bar{y}_{n}}$ the set of photon ToFs in the $n$th pixel, with $\bar{y}_{n} = \sum_{t=1}^{T} y_{n,t}$.
From Eq. \eqref{eq:model_poisson}, it can be shown that the observation model for a given $s_{n}^{p}$ can be expressed as \\
$p(s_{n}^{p}|\boldsymbol{w}_{n},t_{n})$
\begin{equation} \label{form:lik_pstw}
     = \frac{1}{T}\left(1-\sum_{l = 1}^{L} w_{l,n} \right) + \sum_{l = 1}^{L} w_{l,n} \frac{g_{l}(s_{n}^{p} - t_{n})}{G_{l}},
\end{equation}
with $G_{l} = \sum\limits_{t=1}^{T} g_{l}(t - t_{n})$ the integral of the IRF $g_{l}$, which is assumed constant over the admissible values of $t_n$ (such that the IRFs are not cropped).
In \eqref{form:lik_pstw}, the weights $\boldsymbol{w}_{n}=[w_{1,n},\ldots,w_{L,n}]^{\top}$ represent the probabilities of each detected photon to be a signal photons associated with a given wavelength and are given by
\begin{equation} \label{ratio_W}
    w_{l,n} = \frac{r_{n,l}G_{l}}{\sum\limits_{l=1}^{L} r_{n,l}G_{l}  + Tb_{n}}.
\end{equation}
Note that $(1-\sum_{l = 1}^{L} w_{l,n})$ is the probability of a detected photon to be a background photon. The weights are gathered in the $L \times N$ matrix $\boldsymbol{W}=[\boldsymbol{w}_1,\ldots,\boldsymbol{w}_N]$, whose rows are denoted by $\boldsymbol{w}_{l,:}^{\top}$ (as column vectors). 
By construction, the vectors $\boldsymbol{w}_n$ belong to the set 
\begin{equation}
    S_{L} :=  \{\boldsymbol{x}\in \mathds{R}^{L} | \forall l \in \{1,\dots,L\}, x_{l} \geq 0,  \sum\limits_{l = 1}^{L} x_{l} \leq 1 \},
\end{equation}
and we denote by $S_{L}^{N} = \left\{ \boldsymbol{W} | \forall n \in \{1,\ldots,N\}, \boldsymbol{w}_n \in S_{L} \right\}$ the domain of definition of $\boldsymbol{W}$.

Assuming that the ToAs are mutually independent, conditioned on the value of $(\boldsymbol{w}_{n},t_{n})$, the joint likelihood is
\begin{equation} \label{likelihood}
    p(\boldsymbol{S}|\boldsymbol{W},\boldsymbol{t}) = \prod_{n} \prod_{p= 1}^{\bar{y}_{n}} p(s_{n}^{p}|\boldsymbol{w}_{n},t_{n}),
\end{equation}
where the set of ToAs are gathered in $\boldsymbol{S}=\{\boldsymbol{s}_{n}\}_{n=1}^N$.
The next section discusses different prior models associated with the unknown parameters $(\boldsymbol{W},\boldsymbol{t})$ in Eq. \eqref{likelihood}.

\section{Prior Models}
\label{Sec:Priors}

In this work, we address the recovery of $\boldsymbol{W}$ (allowing subsequent estimation of $\boldsymbol{R}$) and $\boldsymbol{t}$ from $\boldsymbol{Y}$ using a Bayesian approach and thus assign prior distributions to these parameters, to incorporate additional information available and regularize the inference problem. Although the depth and reflectivity profiles are expected to be correlated, modeling this correlation is difficult and usually complicates the inference process. To keep the inference process tractable, as in \cite{TV_Yoann_ND,Tobin_SPIRAL,EusipcoYoann,Comp_realData} we assume prior independence between the reflectivity and range profiles. 
Since the estimation quality of $\boldsymbol{t}$ mainly depends on the illumination level (only weakly on the prior model $p(\boldsymbol{t})$, as long as it is informative), a single prior model is used to describe our prior knowledge about $\boldsymbol{t}$, while we consider several prior models for $\boldsymbol{W}$.
        
\subsection{Prior Models for \texorpdfstring{$\boldsymbol{W}$}{Lg}}
\label{subsec:priorW}


\subsubsection{TV-based prior model} 
A classical choice for promoting spatial correlation between parameters in neighboring pixels are Markov random fields (MRFs) \cite{TV_Yoann_ND,Markov_pott,Julian_paper}. The first MRF investigated here is based on a total-variation (TV) regularization \cite{TV_intro_Rudin, TV_Chambolle2004} strategy, which promotes piece-wise constant images, and can be expressed as
\begin{equation} 
\label{prior:TV}
    p(\boldsymbol{W}|\lambda) \propto \exp\left[-\lambda\sum_{l=1}^L\| \boldsymbol{w}_{l,:}^{\top}\|_{TV}\right]\textbf{1}_{S_{L}^{N}}(\boldsymbol{W}),
\end{equation}
where $\textbf{1}_{S_{L}^{N}}(\cdot)$ is the indicator function defined on $S_{L}^{N}$ and $\lambda$ is a user-defined hyperparameter controlling the amount of prior smoothness of the rows of $\boldsymbol{W}$. The $\ell_{1}$-based total variation (TV) regularization of a vectorized image $\boldsymbol{x}$ is defined as $\| \boldsymbol{x}\|_{TV} = \| \nabla_{v}  \boldsymbol{x} \|_{1} + \| \nabla_{h}  \boldsymbol{x} \|_{1}$, where $\nabla_{v}$ (resp. $\nabla_{h}$) is the linear operator computing the discretized vertical (resp. horizontal) finite difference of the image \cite{L1_TV_norm}. 
Note that this prior model is log-concave but not differentiable and that although the rows of $\boldsymbol{W}$ seem a priori independent, they are correlated through the indicator function in Eq. \eqref{prior:TV}. This prior model will be referred as "TV" in this paper.


\subsubsection{Curvature-based prior model}
Another classical MRF is based on the Laplacian operator, 
which imposes an $\ell_{2}$-norm  penalization for the mean (spatial) curvature of images. The resulting model can be expressed as 

\begin{equation} 
\label{prior:Lap}
    f(\boldsymbol{W}|\lambda)\propto  \exp\left[-\frac{\lambda}{2} \sum_{l=1}^L\|\boldsymbol{L}_{a}\boldsymbol{w}_{l,:}^{\top}\|_{2}^{2}\right]\textbf{1}_{S_{L}^{N}}(\boldsymbol{W}),
\end{equation}
where $\boldsymbol{L}_{a}$ is the $(N\times N)$ Laplacian operator \cite{Lap_operator,Lap_operator2}.  
This model is log-concave and differentiable on $S_{L}^{N}$. We will refer to this model as "Lap".
        
\subsubsection{Dirichlet-Based Prior Models} 
\label{DirichletPriors}
The two MRFs presented above promote local spatial correlation but are global models correlating all the image pixels. We also investigate three increasingly informative, yet simpler, prior models that can significantly reduce the computational complexity of the inference process (as will be discussed in Section \ref{Sec:Results}). These models are based on the Dirichlet distribution, which stands as a natural choice to satisfy the constraints induced by $S_{L}^{N}$.

The first and second models are based on
\begin{equation} \label{prior:Dirichlet}
    f(\boldsymbol{W}|\boldsymbol{\beta}) = \prod_{n} \textrm{Dir}( \boldsymbol{w}_n|\boldsymbol{\beta}) {\color{blue}},
\end{equation}
where $\textrm{Dir}(\cdot|\boldsymbol{\beta})$ is the probability density function of the Dirichlet distribution with parameters $\boldsymbol{\beta} = [\beta_{1},\dots,\beta_{L+1}]^{\top}$. Here these parameters are assumed  to take values in $(1,\infty)$ to ensure the strict log-concavity of Eq. \eqref{prior:Dirichlet} with respect to $\textbf{W}$.
Using Eq. \eqref{prior:Dirichlet}, all the vectors $\boldsymbol{w}_n$ share the same hyperparameters, and are a priori independent, conditioned on $\boldsymbol{\beta}$. Moreover the model in Eq. \eqref{prior:Dirichlet} does not promote any spatial correlation and seems consequently well suited when the data is informative enough or when no strong regularization is required. Its main advantage is that it allows the vectors $\boldsymbol{w}_n$ to be updated independently (see Section \ref{Sec:Results}).
Our first model based on Eq. \eqref{prior:Dirichlet} uses a fixed and constant set of hyperparameters, i.e., $\boldsymbol{\beta} = \kappa \textbf{1}_{L+1}$, where $\textbf{1}_{L+1}$ is the $(L+1)$ column vector of ones. In practice $\kappa$ is set to $\kappa = 1.01$ to obtain a log-concave, weakly informative prior model. This first model is referred to as "W-Dirichlet" (weak Dirichlet-based model) in the remainder of the paper. 

The second prior model embeds Eq. \eqref{prior:Dirichlet} in a hierarchical model, i.e., considers $\boldsymbol{\beta}$ is an unknown parameter vector which is assigned as prior model a product of $(L+1)$ independent and identical truncated exponential distributions with fixed hyperparameter $\theta$, that is 
\begin{equation} \label{hyperprior}
    f(\boldsymbol{\beta}|\theta) \propto\prod_{l=1}^{L+1} \theta e^{-\theta\beta_{l}} \textbf{1}_{(1,\infty)}(\boldsymbol{\beta}).
\end{equation}
In this case, $\boldsymbol{\beta}$ is estimated together with $\boldsymbol{W}$ and the truncation is introduced to ensure the log-concavity of \eqref{prior:Dirichlet} with respect to $\boldsymbol{W}$.  In all the results presented in Section \ref{Sec:Results} we fixed $\theta=1/4$. This model is referred to as "G-Dirichlet" (global Dirichlet-based model with unknown $\boldsymbol{\beta}$).

The third Dirichlet-based model is similar to the G-Dirichlet model but is applied to groups of pixels instead of the whole image \cite{PatchPrior}. Let us assume that the image pixels are clustered in $C$ distinct and known groups, each group presenting specific spectral profiles. The procedure to form these groups, which are generally unknown will be discussed in Section \ref{subsec:comp_cons}. Let $I_c$ be the set of pixel indices in the $c$th group. The joint prior model for $\boldsymbol{W}$ can be written as 
\begin{eqnarray}
\label{prior:CDir}
    f(\boldsymbol{W}|\boldsymbol{\beta}_1,\ldots,\boldsymbol{\beta}_C) = \prod_{c=1}^C\prod_{n\in I_c} \textrm{Dir}( \boldsymbol{w}_n|\boldsymbol{\beta}_c),
\end{eqnarray}
where $\boldsymbol{\beta}_c$ is a group-dependent vector of hyperparameters. In a similar fashion to the G-Dirichlet, the vectors $\{\boldsymbol{\beta}_c\}_c$ are assumed unknown, a priori independent and are assigned prior models as in Eq. \eqref{hyperprior}. As for the first two Dirichlet-based models, this model does not explicitly account for local correlation between pixels (the vectors $\boldsymbol{w}_n$ are a priori mutually independent for a fixed clustering and given values of $\boldsymbol{\beta}_1,\ldots,\boldsymbol{\beta}_C$) but (local and non-local) correlation can be implicitly introduced when forming the clusters (see Section \ref{subsec:comp_cons}).
We will refer to this model as "C-Dirichlet" (cluster-based Dirichlet-based model).

In Section \ref{Sec:Strategy}, we denote by $\bPhi$ the set of unknown hyperparameters involved in the prior model of $\boldsymbol{W}$, i.e., $\bPhi=\emptyset$ for Eqs. \eqref{prior:TV} and \eqref{prior:Lap} and W-Dirichlet, $\bPhi=\boldsymbol{\beta}$ for G-Dirichlet and $\bPhi=\{\boldsymbol{\beta}_1,\ldots,\boldsymbol{\beta}_C\}$ for C-Dirichlet.

\subsection{Prior Model for the Range Profile}
 \label{depth_reg}
Recent single-photon avalanche diode (SPAD) detectors \cite{SPAD_det} offer a timing resolution of several picoseconds which allows sub-centimeter range resolution. Thus, it makes sense to assume that the elements of $\boldsymbol{t}$ are defined on the discrete grid with the same resolution, as in \cite{model_Bay1, TV_Yoann_1D}, and take values in $\{t_{min},\ldots,t_{max}\}$ such that $1<t_{min}<t_{max}<T$. Using discrete depths also simplifies the estimation of $\boldsymbol{t}$. The bounds $(t_{min},t_{max})$ ensure that the integrals $\{G_l\}_l$ are indeed constant, irrespective of the position of the objects. Since real scenes are often composed of distinct surfaces, it is reasonable to define a prior model for $\boldsymbol{t}$ which preserves sharp edges. As in \cite{Tobin_SPIRAL, TV_Yoann_ND}, the following TV-based MRF is used in this work
\begin{equation}
\label{eq:prior_t}
    p(\boldsymbol{t}|\epsilon) = \exp\left[ -\epsilon\| \boldsymbol{t}\|_{TV}\right], 
\end{equation}
where the fixed (user-defined) hyperparameter $\epsilon$, which mostly depends on the range resolution of the single-photon detector, determines the level of correlation between the depth parameters of neighboring pixels. In Section \ref{Sec:Results}, we set $\epsilon=0.05$ for all the experiments.


\section{Estimation Strategy} \label{Sec:Strategy}

In Section \ref{Sec:Priors}, we defined joint prior models for $(\boldsymbol{W},\boldsymbol{t})$ or $(\boldsymbol{W},\boldsymbol{t},\bPhi)$ (for the G-Dirichlet and C-Dirichlet models). In this section, we primarily present the estimation strategy to estimate $(\boldsymbol{W},\boldsymbol{t})$ (and subsequently $\boldsymbol{R}$) for ease of understanding. Cases where $\bPhi$ is not empty will be only briefly discussed in Section \ref{Sec:PHI} as $\bPhi$ can be updated together with $\boldsymbol{W}$. Note that the fixed hyper-parameters (e.g., $\epsilon$ and $\lambda$ for the models in Eqs. \eqref{prior:TV}-\eqref{prior:Lap}) are omitted for brevity in the equations of the remainder of this paper. 
Using the Bayes rule, the joint posterior distribution of $(\boldsymbol{W},\boldsymbol{t})$ can be expressed as 
\begin{equation} \label{Posterior}
    p(\boldsymbol{W},\boldsymbol{t}|\boldsymbol{S},\bPhi) \propto  p(\boldsymbol{S}|\boldsymbol{W},\boldsymbol{t})f(\boldsymbol{W})p(\boldsymbol{t}).
\end{equation}
However, as mentioned previously, the joint estimation of $\boldsymbol{W}$ and $\boldsymbol{t}$ is challenging, mainly due to the multimodal nature of $p(\boldsymbol{S}|\boldsymbol{W},\boldsymbol{t})$. To circumvent this issue, the joint estimation problem is decomposed into two successive problems where we first consider $\boldsymbol{t}$ as a nuisance vector to be marginalized to estimate $\boldsymbol{W}$ via  marginal maximum a posteriori (MMAP) estimation.
Once $\boldsymbol{W}$ is estimated, the depth profile can then be inferred from the posterior distribution $p(\boldsymbol{t}|\widehat{\boldsymbol{W}}_{MMAP},\boldsymbol{S})$, e.g., via MMAP estimation, conditioned on the estimated proportions $\widehat{\boldsymbol{W}}_{MMAP}$.

\subsection{Estimation of mixture fractions} \label{sec:EM_part1}
The estimation of $\boldsymbol{W}$ is performed using
\begin{eqnarray}
\widehat{\boldsymbol{W}}_{MMAP} =\underset{\boldsymbol{W}}{\textrm{argmax}}~~f(\boldsymbol{W}|\boldsymbol{S}),
\end{eqnarray}
which can be performed using an EM-based algorithm. A typical iteration of the classical EM algorithm \cite{EM1977} is decomposed into two steps. Let $\boldsymbol{W}^{(i)}$ be the current estimate of $\boldsymbol{W}$. The first step consists of computing
\begin{eqnarray}
\label{eq:classical_E}
Q(\boldsymbol{W}|\boldsymbol{W}^{(i)})=\textrm{E}_{p(\boldsymbol{t}|\boldsymbol{W}^{(i)},\boldsymbol{S})}\left[ \log p(\boldsymbol{W},\boldsymbol{t}|\boldsymbol{S}) \right], 
\end{eqnarray}
while the second step updates the estimate of $\boldsymbol{W}$ via
\begin{eqnarray}
\label{eq:classical_M}
\boldsymbol{W}^{(i+1)}=\underset{\boldsymbol{W} \in S_{L}^{N}}{\textrm{argmax}}~~ Q(\boldsymbol{W}|\boldsymbol{W}^{(i)}).
\end{eqnarray}
Unfortunately, the first step in \eqref{eq:classical_E} cannot be computed directly as the expectation $\textrm{E}_{p(\boldsymbol{t}|\boldsymbol{W}^{(i)},\boldsymbol{S})}\left[\boldsymbol{t}\right]$
is intractable because of the MRF prior distribution assigned to $\boldsymbol{t}$ in \eqref{eq:prior_t}.
In such cases, a classical approach is to resort to stochastic EM (SEM) algorithms \cite{EM_Book,celeux1996stochastic}. 
In this work, to keep the overall computational cost of the method low, we adopt the strategy detailed in \cite{Celeux2003}. More precisely, we replace $p(\boldsymbol{t}|\boldsymbol{W}^{(i)},\boldsymbol{S})$ in \eqref{eq:classical_E} by an approximating distribution such that the expectation becomes tractable. In a similar fashion to \cite{MC_neighboor}, the approximating distribution is constructed by generating a random sample $\tilde{\boldsymbol{t}}$ from $p(\boldsymbol{t}|\boldsymbol{W}^{(i)},\boldsymbol{S})$ and we use the approximation
\begin{eqnarray} \label{eq:equiv_prior_samp}
p(\boldsymbol{t}|\boldsymbol{W}^{(i)},\boldsymbol{S}) \approx \tilde{p}(\boldsymbol{t}|\tilde{\boldsymbol{t}},\boldsymbol{W}^{(i)},\boldsymbol{S}),
\end{eqnarray}
which can significantly simplify the computation of expectations. Indeed, we use 
\begin{eqnarray}
\tilde{p}(\boldsymbol{t}|\tilde{\boldsymbol{t}},\boldsymbol{W}^{(i)},\boldsymbol{S})=\prod_{n=1}^N p(t_n|\tilde{\boldsymbol{t}}_{\backslash n},\boldsymbol{W}^{(i)},\boldsymbol{S}),
\end{eqnarray}
where $\tilde{\boldsymbol{t}}_{\backslash n}$ is the vector $\tilde{\boldsymbol{t}}$ whose $n$th element has been removed and $p(\cdot|\tilde{\boldsymbol{t}}_{\backslash n},\boldsymbol{W}^{(i)},\boldsymbol{S})$ is the conditional distribution of the $n$th element of $\tilde{\boldsymbol{t}}$ associated with the joint distribution $p(\tilde{\boldsymbol{t}}|\boldsymbol{W}^{(i)},\boldsymbol{S})$. The approximation thus reduces to a product of $N$ independent distributions and expectations become easier to compute.

Instead of computing \eqref{eq:classical_E}-\eqref{eq:classical_M}, the two EM steps become
\begin{eqnarray}
\label{eq:modified_E}
\widetilde{Q}(\boldsymbol{W}|\boldsymbol{W}^{(i)})=\textrm{E}_{\tilde{p}(\boldsymbol{t}|\tilde{\boldsymbol{t}},\boldsymbol{W}^{(i)},\boldsymbol{S})}\left[ \log p(\boldsymbol{W},\boldsymbol{t}|\boldsymbol{S}) \right], 
\end{eqnarray}
and
\begin{eqnarray}
\label{eq:modified_M}
\boldsymbol{W}^{(i+1)}=\underset{\boldsymbol{W} \in S_{L}^{N}}{\textrm{argmax}}~~ \widetilde{Q}(\boldsymbol{W}|\boldsymbol{W}^{(i)}).
\end{eqnarray}

Defining $\tilde{p}_{n,k}=p(t_n=k|\tilde{\boldsymbol{t}}_{\backslash n},\boldsymbol{W}^{(i)},\boldsymbol{S}), \forall (n,k)$, we obtain
\begin{eqnarray}
\widetilde{Q}(\boldsymbol{W}|\boldsymbol{W}^{(i)})=\widetilde{Q}_1(\boldsymbol{W}|\boldsymbol{W}^{(i)})+\widetilde{Q}_2(\boldsymbol{W}|\boldsymbol{W}^{(i)}) + C_1,
\end{eqnarray}
where $C_1$ is a constant which does not depend on $\boldsymbol{W}$, $\widetilde{Q}_1(\boldsymbol{W}|\boldsymbol{W}^{(i)})= \log \left( f(\boldsymbol{W})\right)$ and 
\begin{eqnarray}
\widetilde{Q}_2(\boldsymbol{W}|\boldsymbol{W}^{(i)}) = \sum_{n,k} \tilde{p}_{n,k} \sum_{p= 1}^{\bar{y}_{n}} \log \left (p(s_{n}^{p}|\boldsymbol{w}_{n},t_{n}=k)\right ).
\end{eqnarray}

The likelihood $p(s_{n}^{p}|\boldsymbol{w}_{n},t_{n})$ in Eq. \eqref{form:lik_pstw} is concave with respect to (w.r.t.) $\boldsymbol{W}$, which results in $\widetilde{Q}(\boldsymbol{W}|\boldsymbol{W}^{(i)})$ being a concave function w.r.t. $\boldsymbol{W}\in S_{L}^{N}$. Thus, the solution in \eqref{eq:modified_M} can be obtained using convex optimization methods. This will be discussed further in Section \ref{subsec:comp_cons}.


\subsection{Estimation of the hyper-parameters} \label{Sec:PHI}
The method proposed can be easily extended to estimate the hyper-parameters in $\bPhi$ associated with $\boldsymbol{W}$. In that case, Eqs. \eqref{eq:modified_E} and \eqref{eq:modified_M} become\\
$\widetilde{Q}(\boldsymbol{W},\bPhi|\boldsymbol{W}^{(i)},\bPhi^{(i)})$
\begin{eqnarray}
\label{eq:modified_E_2}
=\textrm{E}_{\tilde{p}(\boldsymbol{t}|\tilde{\boldsymbol{t}},\boldsymbol{W}^{(i)},\boldsymbol{S},\bPhi^{(i)})}\left[ \log p(\boldsymbol{W},\boldsymbol{t},\bPhi|\boldsymbol{S}) \right], 
\end{eqnarray}
and
\begin{eqnarray}
\label{eq:modified_M_2}
(\boldsymbol{W}^{(i+1)},\bPhi^{(i+1)})=\underset{\boldsymbol{W} \in S_{L}^{N},\bPhi}{\textrm{argmax}}~~ \widetilde{Q}(\boldsymbol{W},\bPhi|\boldsymbol{W}^{(i)},\bPhi^{(i)}), 
\end{eqnarray}
where 
\begin{eqnarray}
p(\boldsymbol{W},\boldsymbol{t},\bPhi|\boldsymbol{S}) \propto  p(\boldsymbol{S}|\boldsymbol{W},\boldsymbol{t})p(\boldsymbol{t})p(\boldsymbol{W|\bPhi})p(\bPhi), 
\end{eqnarray}
and $p(\bPhi)$ consists of a product of truncated exponential distributions (see Section \ref{DirichletPriors}).
Sampling the auxiliary variable $\tilde{\boldsymbol{t}}$ needed in \eqref{eq:modified_E_2} can be achieved as when $\bPhi=\emptyset$. Optimizing \eqref{eq:modified_M_2} globally is challenging as the problem is not convex and alternating optimization w.r.t. $\boldsymbol{W}$ and $\bPhi$ can be adopted to ensure local convergence. Instead, to reduce the computational complexity, we simply solve 
\begin{eqnarray}
\label{eq:modified_M_3}
\left\{
    \begin{array}{l}
        \boldsymbol{W}^{(i+1)}=\underset{\boldsymbol{W} \in S_{L}^{N}}{\textrm{argmax}}~~ \widetilde{Q}(\boldsymbol{W},\bPhi^{(i)}|\boldsymbol{W}^{(i)},\bPhi^{(i)})\\
        \bPhi^{(i+1)}=\underset{\bPhi}{\textrm{argmax}}~~ \widetilde{Q}(\boldsymbol{W}^{(i+1)},\bPhi|\boldsymbol{W}^{(i)},\bPhi^{(i)})
    \end{array}
\right.,
\end{eqnarray}
leading to a generalized EM algorithm \cite{EM_Book}. Both lines in Eq. \eqref{eq:modified_M_3} can be solved using convex optimization as Eq. \eqref{eq:modified_M}.

The resulting algorithm is summarized in Algo. \ref{algo:algo1}. Due to the stochastic nature of the EM-based method used, i.e., due to the sampling step (Line 5 in Algo. \ref{algo:algo1}), the proposed iterative procedure does not converge exactly to the MMAP estimator of $\boldsymbol{W}$ but after some iterations, referred to as burn-in iterations, the generated $\boldsymbol{W}^{(i)}$ oscillate around it. The number of burn-in iterations $N_{bi}$ and the total number of iterations $N_{iter}$ are user-defined parameters. For all the simulations results presented in Section \ref{Sec:Results} however, the convergence of the algorithm is fast (less than 5-10 iterations) and the oscillations are quite small. This can be monitored using the relative error between successive $\boldsymbol{W}^{(i)}$. Thus, we stop the burn-in when the error becomes smaller than a fixed threshold $d_{\epsilon} = 10^{-10}$. Since the oscillations are small in practice, we used $N_{iter}=N_{bi}+5$. After $N_{iter}$ iterations, we finally use the average of the last $N_{\textrm{iter}}-N_{\textrm{bi}}$ generated values of $\boldsymbol{W}$ as our estimate $\widehat{\boldsymbol{W}}_{MMAP}$ (Line 14 in Algo. \ref{algo:algo1}).

\begin{algogo}{EM-based estimation of $\boldsymbol{W}$}
\label{algo:algo1}
 \begin{algorithmic}[1]
			\STATE \underline{Fixed input parameters:} $\epsilon,\lambda$, number of burn-in iterations $N_{\textrm{bi}}$, total number of iterations $N_{\textrm{iter}}>N_{\textrm{bi}}$.
			\STATE \underline{Initialization ($k=0$)}
			\STATE Set $\boldsymbol{W}^{(0)}$ (and $\bPhi^{(0)}$)
			\FOR{$i=0,\ldots N_{\textrm{iter}}$}
			\STATE Sample $\tilde{\boldsymbol{t}} \sim p(\boldsymbol{t}|\boldsymbol{W}^{(i)},\boldsymbol{S},\bPhi^{(i)})$.
			\STATE Compute $\tilde{p}_{n,k}=p(t_n=k|\tilde{\boldsymbol{t}}_{\backslash n},\boldsymbol{W}^{(i)},\boldsymbol{S},\bPhi^{(i)}), \forall (n,k)$
			\STATE Compute $\boldsymbol{W}^{(i+1)}$ using Eq. \eqref{eq:modified_M} (or \eqref{eq:modified_M_3}).
			\IF {$\bPhi^{(i)}\neq \emptyset$}
			\STATE Compute $\bPhi^{(i+1)}$ using Eq. \eqref{eq:modified_M_3}.
			\ELSE
			\STATE Set $\bPhi^{(i+1)}=\bPhi^{(i)}$.
			\ENDIF
\ENDFOR
		\STATE Set $\widehat{\boldsymbol{W}}_{MMAP}=1/(N_{\textrm{iter}}-N_{\textrm{bi}})\sum_{i=N_{\textrm{bi}}+1}^{N_{\textrm{iter}}} \boldsymbol{W}^{(i)}$
\end{algorithmic}
\end{algogo} 

\subsection{Estimation of the spectral responses}
\label{Sec:R}

The algorithm presented in the previous section estimates $\boldsymbol{W}$ but does not provide directly an estimate of $\boldsymbol{R}$. It can be seen from \eqref{eq:model_poisson_MSL} that
\begin{eqnarray}
E_{p\left(\boldsymbol{y}_{n}|\boldsymbol{r}_{n},t_{n},b_{n}\right)}[\bar{y}_{n}] = \sum_{l=1}^{L} r_{n,l}G_{l}  + Tb_{n},
\end{eqnarray} i.e., that $r_{n,l} = \frac{w_{n,l} E[\boldsymbol{y}_{n}]}{G_{l}}$ using \eqref{ratio_W}. While it is possible to estimate $r_{n,l}$ using $\hat{w}_{n,l} \bar{y}_{n}/G_{l}$, this estimate, which assumes 
$\bar{y}_n\approx E[\bar{y}_{n}]$, is in practice too noisy, especially in the photon-starved regime. Thus, we propose to denoise the image $\bar{\boldsymbol{y}}=\{\bar{y}_n\}_n$ to get a better estimate $\hat{\boldsymbol{y}}=\{\hat{y}_n\}_n$, leading to
\begin{eqnarray} \label{eq:estim_R}
\hat{r}_{n,l} = \frac{\hat{w}_{n,l} \hat{y}_{n}}{G_{l}}.
\end{eqnarray}
Using the fact that $\bar{\boldsymbol{y}}$ is corrupted by Poisson noise, $\hat{\boldsymbol{y}}$ is obtained by denoising the image $\bar{\boldsymbol{y}}$ using a variant of the BM3D algorithm \cite{BM3D_Gaussian}, i.e., the method developed in \cite{BM3D_Poisson} to account for Poisson noise. While other denoising algorithms could be used, this denoiser provides state-of-the-art reconstruction results with a reduced computational time and limited user supervision. Here, we used the default parameter settings of the code available at \url{http://www.cs.tut.fi/~foi/invansc/}.

\subsection{Estimation of the depth profile}
\label{subsec:estim_t}
Following the computation of $\widehat{\boldsymbol{W}}_{MMAP}$ via the EM-based algorithm, 
the depth profile is finally estimated using
\begin{equation}\label{posterior_T}
    \hat{t}_{n,MMAP} = \underset{t_{n}}{\textrm{argmax}} \sum_{\boldsymbol{t}_{\setminus n}} p(\boldsymbol{t}|\widehat{\boldsymbol{W}}_{MMAP},\boldsymbol{S}), \forall n
\end{equation}
These estimates are obtained via Monte Carlo simulation from $p(\boldsymbol{t}|\widehat{\boldsymbol{W}}_{MMAP},\boldsymbol{S})$, which can be performed efficiently using a 2-step Gibbs sampler, exploiting the MRF structure of the TV-based prior model \eqref{eq:prior_t}, as in \cite{TV_Yoann_1D}.
In all the experiments carried out and presented in Section \ref{Sec:Results}, the Gibbs sampler used to estimate the depth profile was stopped after 300 iterations, with a burn-in period of 50 iterations.


\subsection{Computational considerations}
\label{subsec:comp_cons}
\underline{Mazimization step:} although a thorough study of algorithms to solve \eqref{eq:modified_M} is out of scope of this paper, we discuss here two different approaches that have been used depending on the prior model for $\boldsymbol{W}$. With the prior models \eqref{prior:TV} or \eqref{prior:Lap}, we used the classical alternating direction method of multipliers (ADMM) \cite{BoydBook} as it can easily handle the constraints on $\boldsymbol{W}$ as well as smooth or non-smooth prior models. The same ADMM stopping criterion as in \cite{BoydBook} has been used in all our experiments. Since the splitting schemes (e.g., the number of splitting variables) varies, the different ADMM schemes implemented are not further detailed here. Note however that some steps (e.g., involving $\widetilde{Q}_1(\boldsymbol{W}|\boldsymbol{W}^{(i)})$) in the ADMM sequential approach cannot be computed analytically and require optimization sub-routines. In this case, where the cost function is differentiable, we adopted a line search approach. 
Since a variable-splitting approach is used, the overall computational cost of the method can be significant, as it requires storing several matrices (the splitting variables) that have the same size as $\boldsymbol{W}$. Using the Dirichlet-based models, the optimization in \eqref{eq:modified_M} can be performed pixel-wise and $\widetilde{Q}(\boldsymbol{W}|\boldsymbol{W}^{(i)})$ is twice differentiable and strictly concave. Thus we can use second-order methods, e.g., a simple Newton-Raphson (NR) algorithm 
to solve \eqref{eq:modified_M} efficiently. When $\bPhi\neq\emptyset$, the second line of Eq. \eqref{eq:modified_M_3} is solved iteratively using a component wise NR algorithm, where the elements of $\bPhi$ are updated sequentially until convergence. As will be seen in Section \ref{Sec:Results}, this makes Dirichlet-based models computationally more attractive than MRF-based models.

\underline{Sampling the auxiliary variable $\boldsymbol{t}$:} the sampling step in Line 5 of Algo. \ref{algo:algo1} requires sampling from a discrete MRF, which can be done in a similar fashion to the final estimation of $\boldsymbol{t}$ in Section \ref{subsec:estim_t}. However, using a Gibbs sampling approach with a random initialization would yield a long burn-in period to obtain a sample from $p(\boldsymbol{t}|\boldsymbol{W}^{(i)},\boldsymbol{S},\bPhi^{(i)})$. Instead, we use the value of $\tilde{\boldsymbol{t}}$ generated at the previous iteration to hot-start the Gibbs sampler and only iterate a reduced number of times (1 to 2 in practice) to generate a sample.

\underline{MRF hyperparameter $\lambda$:} the dynamic range of $\boldsymbol{W}$ decreases in the presence of high background and $\lambda$ in \eqref{prior:TV} and Eq. \eqref{prior:Lap} should be adjusted accordingly if prior information about the signal to background ratio is available. In Section \ref{Sec:Results} however, we do not use such information and set $\lambda = 10$ in \eqref{prior:TV} and Eq. \eqref{prior:Lap} for all the experiments as this value leads to satisfactory results on average, irrespective of the level of background in the range considered. Note that the level of user-supervision required for fine-tuning $\lambda$ is one of the motivations for using the C-Dirichlet model, whose parameters are more robust to illumination conditions.

\underline{Pixel clustering for the C-Dirichlet model:} the C-Dirichlet model defined in Section \ref{Sec:Priors} relies on a user-defined clustering of the pixels, which is difficult to set beforehand. To define the $C$ groups of pixels, we first run Algo. \ref{algo:algo1} for a few iterations using the W-Dirichlet model to get a coarse initial guess $\bar{\boldsymbol{W}}_0$ of $\boldsymbol{W}$ which is used to create the clusters. 
While direct clustering of the $N$ $L$-dimensional vectors in $\bar{\boldsymbol{W}}_0$ is possible, it is in practice prone to noise and patch-based clustering is preferred here.
More precisely, we first construct from $\bar{\boldsymbol{W}}_0$ a set of $N$ 3D patches of size $N_p\times N_p \times L$ (one patch per pixel). 
The patches are then vectorized and clustered into $C$ groups via k-means clustering with the $\ell_{2}$ distance as similarity measure between patches \cite{benjamini2016white}. 
In all the results presented in Section \ref{Sec:Results}, we used $C=7$ classes of $(3\times 3)$ pixels patches. Note however that different clustering strategies, e.g.,  \cite{chuang2006fuzzy,zhao2016spectral,camps2013advances} could be used instead of k-means.

\section{Results} \label{Sec:Results}

We first assess the performance of the proposed approach with the different prior models for $\boldsymbol{W}$ using synthetic SBL data, i.e., with $L=1$ in Section \ref{res:syntthetic_SBL} and then apply our approach on real SW-MSL data in Section \ref{res:RD_SW-MSL} with a reduced number of prior models.

\subsection{Analysis of synthetic SBL data}
\label{res:syntthetic_SBL}
 
\begin{figure}[htb]
 \begin{center}
\resizebox{88mm}{!}{\includegraphics{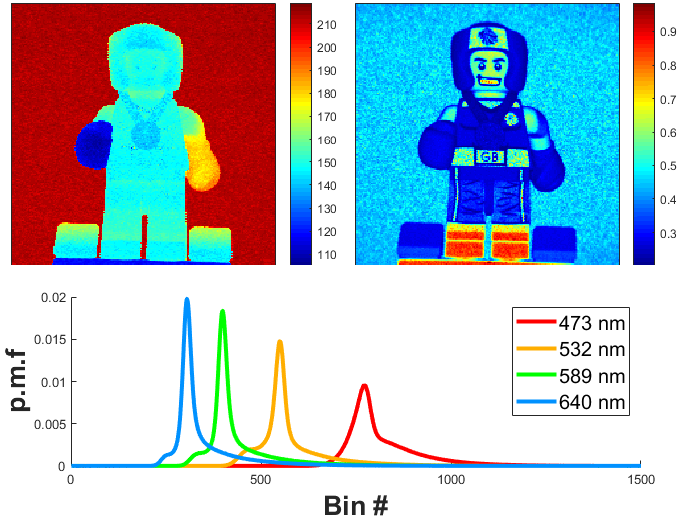}}
 \end{center}
 \caption{Top: depth (left) and reflectivity profiles used to simulate SBL data. Bottom: normalised IRFs from \cite{Tobin_SPIRAL}, corresponding to the probability mass functions (p.m.f) of the photons ToAs associated with the wavelengths $[473, 532, 589, 640]$nm. The orange IRF is that used in Section \ref{res:syntthetic_SBL}.}
 \label{Figurine}
\end{figure}

In this section, we generated synthetic data using the $N = 200\times 200$ pixels depth and reflectivity profiles, depicted in Fig. \ref{Figurine} (top), which are actual depth and reflectivity (at $532$nm) profiles obtained in \cite{Tobin_SPIRAL} and \cite{Comp_realData}. The associated IRF $g(.)$ shown in Fig. \ref{Figurine} (bottom) in orange ($532$nm) was obtained during the calibration of the imaging system.
The number of temporal bins was set to $T = 1500$ (bin width of $2$ps) and a spatially constant background $\boldsymbol{b} = \{b_{n}\}_{n}$ was added to all of the waveforms. For all of the experiments presented in this paper, we set the admissible temporal positions to $[t_{min},t_{max}]=[301,T-600]$ to ensure the integrals of the IRFs remain constant over the admissible object range.

To control the quality of the generated data, we introduce as in \cite{EusipcoYoann}, two parameters ($\alpha$ and $\gamma$) controlling the mean signal detection and the background levels. More precisely, datasets have been generated using Eq. \eqref{eq:model_poisson}, such that the average number of signal and background photons in the $n$th pixel are $\alpha r_{n} G_{1}$ and $b_{n} = \alpha \gamma T$, respectively.
The parameter $\alpha$ is chosen in $\{25,100\}$ and $\gamma$ takes values in $\{0.01,0.05,0.1,0.3,0.5,1,5\}/T$. This leads to average counts per pixel from  $10.3$ (for $(\alpha,\gamma)=(25,0.01/T)$) to $542.7$ (for $(\alpha,\gamma)=(100,5/T)$). In the latter case, about $500$ of these photons arise from the background. With this parametrization, and average signal to background ratio (SBR)
\begin{equation*}
    \text{SBR} = \frac{1}{N}\sum_{n=1}^{N} \frac{1}{T b_{n}}\sum\limits_{l=1}^{L} r_{n,l} G_{l},
\end{equation*}
can be expressed as $\text{SBR} = \frac{1}{NT \gamma}\sum_{n=1}^{N} \sum_{l=1}^{L} r_{n,l} G_{l}$,
and it ranges from  $0.086$ (for $\gamma=5/T$) to $42$ (for $\gamma=0.01/T$).

The proposed approach is compared in the SBL case with two approaches designed to reconstruct 3D surfaces assuming exactly one visible surface per pixel:

\underline{Cross-correlation approach:} denoted as "Xcorr", it corresponds to the classical matched-filter \cite{McCarthy_LogMatchFilt}. The depth profile is estimated by maximizing the correlation between the IRF and each histogram $\boldsymbol{y}_{n}$. These estimates are then used in Eq. \eqref{form:lik_pstw} to estimate $\boldsymbol{W}$ via maximum likelihood, and then $\boldsymbol{R}$ using the method described in Section \ref{Sec:R}. This strategy yields better reflectivity estimates than the standard estimation of $(\boldsymbol{R},\boldsymbol{b})$ from Eq. \eqref{form:lik_pstw}, conditioned on the estimated depth profile.

\underline{Photon unmixing algorithm \cite{model_Bay1}:} Pixel patches of size $5\times 5$ have been used to locally improve the SBR and the probability threshold has been set to $\tau_{FA}=0.05$ to censor background photons before parameter estimation (see \cite{model_Bay1} for additional details). These values have been optimized manually to obtain the best results on average for each $(\alpha,\gamma)$, according to the metric in \eqref{MSE_Multi}. This method assumes that the SBR is known.

Note that the method proposed in \cite{EusipcoYoann} coincides with our Lap model for $L=1$, i.e., when the Dirichlet distribution becomes a beta distribution, and this method is thus not included above.

The reflectivity estimation is assessed using the mean squared error
\begin{equation} \label{MSE_Multi}
    \text{MSE} = \frac{1}{N} \sum\limits_{n=1}^{N}  \|\boldsymbol{r}_{n} - \hat{\boldsymbol{r}}_{n}\|_{2}^2,
\end{equation}
where $\boldsymbol{r}_n$ (resp. $\hat{\boldsymbol{r}}_n$) is the actual (resp. estimated) spectral response of the $n$th pixel.
\begin{figure}[htb]
    \begin{center}
\resizebox{88mm}{!}{
    \includegraphics{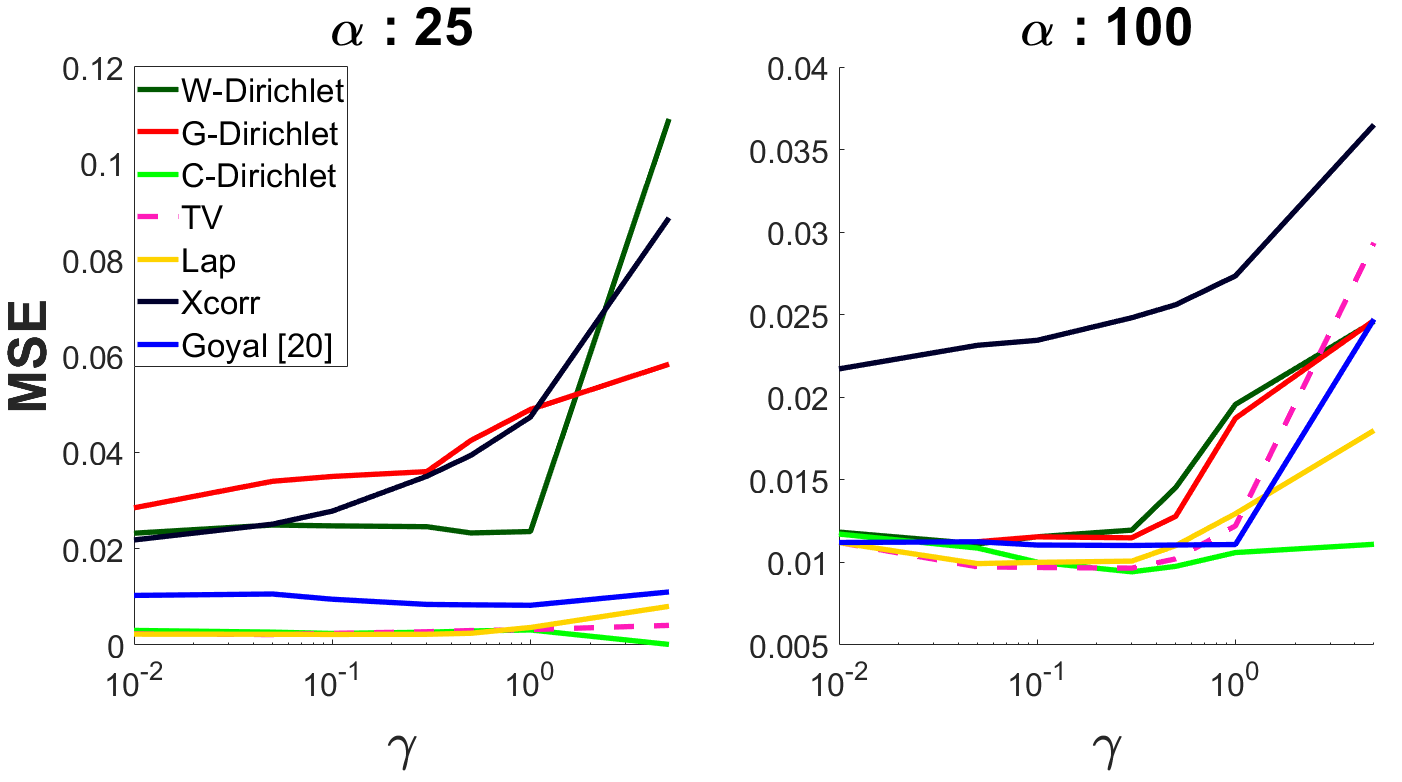}}
    \end{center}
    \caption{MSE of the reflectivity (averaged over 3 realizations) obtained with the different competing methods for $L=1$, as function of $\gamma$, with $\alpha=25$ (left) and $\alpha=100$ (right).\label{fig:MAE_R_1D.png}}
\end{figure}
The MSEs obtained with the different methods for $\alpha=25$ and $\alpha=100$ as a function of $\gamma$ are shown in Fig. \ref{fig:MAE_R_1D.png}. In the low illumination scenario ($\alpha = 25$), this figure shows that the TV, Lap and C-Dirichlet models perform similarly well and that the C-Dirichlet model provides the most accurate estimate for high values of $\gamma$. The W-Dirichlet and G-Dirichlet models provide the least accurate reflectivity reconstruction. The results obtained with the Xcorr model are close to that of W-Dirichlet, explained by the lack of strong regularization.
We obtained similar results in the high illumination scenario ($\alpha = 100$), where only the TV, Lap and C-Dirichlet models allow more accurate reconstruction than the method from \cite{model_Bay1} for low background levels ($\gamma<0.5$), and only the C-Dirichlet provides better reconstruction in term of the MSE than the method of Goyal \cite{model_Bay1}.

\begin{figure}[htb]
    \begin{center}
\resizebox{88mm}{!}{
    \includegraphics{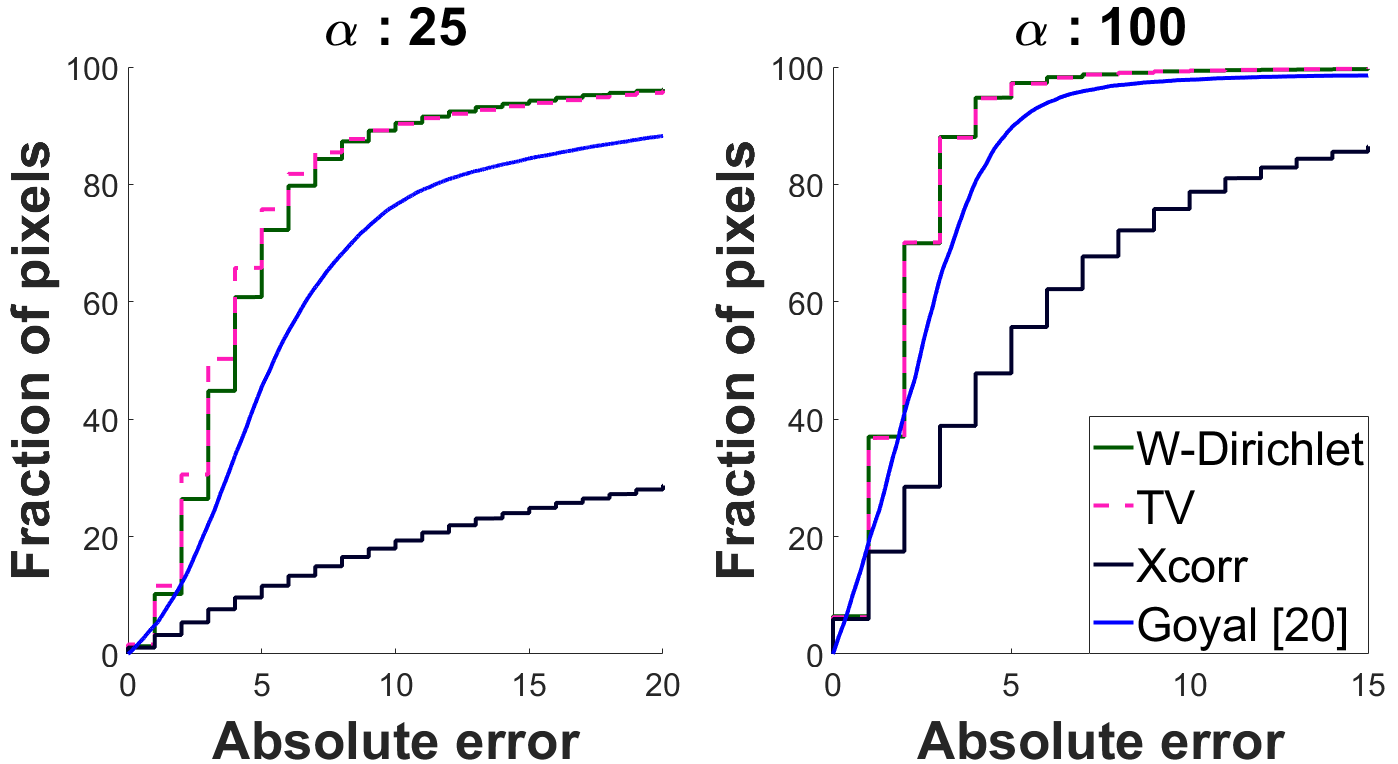}}
    \end{center}
    \caption{CDFs of the depth absolute error obtained using the competing methods for data generated with $L=1$, $\gamma=5$ and $\alpha=25$ (left) and $\alpha=100$ (right). 
    \label{fig:CDF_1D}
    }
\end{figure}

The ranging performance is quantified using the cumulative density function (CDF) of the depth absolute error $h_{n} = |t_{n} - \hat{t}_{n}|, \forall n$, where $\hat{t}_{n}$ is the estimate of the actual depth parameter $t_n$ in the $n$th pixel. 
The depth error CDFs obtained with $\alpha \in \{25,100\}$ are depicted in Fig. \ref{fig:CDF_1D}. Note that TV and Lap perform similarly and that the Dirichlet-based priors also perform similarly. Thus the results obtained with Lap, G-Dirichlet and C-Dirichlet are omitted here.
In the low illumination scenario, Xcorr provides the poorest estimation, whereas TV, similarly to W-Dirichlet, achieves the most accurate. Note that among the prior models we proposed, W-Dirichlet and TV provide respectively the least and most accurate estimates and both perform better than the method from \cite{model_Bay1} for $\alpha=25$. 
With $\alpha=100$, all the methods (except Xcorr) perform similarly.
Interestingly, the choice of the prior model for $\boldsymbol{W}$ has in general a limited impact on the quality of $\hat{\boldsymbol{t}}$, which is more affected by the (lack of) depth regularization.

\begin{table}[htb]
\begin{center}
 \begin{tabular}{|c|c|c|c|c|c|c|}
 \hline
  \multirow{2}{*}{$(\alpha,\gamma)$} & Counts / & \multirow{2}{*}{Method} & \multirow{2}{*}{$N_{iter}$} & Time /& Total \\
   & SBR & & & iter & time \\
 \hline
  \multirow{3}{*}{(25,0.01)} & \multirow{3}{*}{10.3/42} & W-Dirichlet & 12 & 7s & 93s \\
  & & C-Dirichlet & 12 & 8s & 95s \\
  & & TV & 12 & 10s & 127s \\ 
   \hline
  \multirow{3}{*}{(25,5)} & \multirow{3}{*}{135.6/0.086} & W-Dirichlet & 12 & 12s & 155s \\
  & & C-Dirichlet & 11 & 11s & 128s \\
  & & TV & 12 & 14s & 157s \\ 
   \hline
  \multirow{3}{*}{(100,0.01)} & \multirow{3}{*}{43.8/42} & W-Dirichlet & 11 & 17s & 190s \\
  & & C-Dirichlet & 12 & 16s & 197s \\
  & & TV & 12 & 19s & 230s \\ 
   \hline
  \multirow{3}{*}{(100,5)} & \multirow{3}{*}{542.7/0.086} & W-Dirichlet & 12 & 27s & 300s \\
  & & C-Dirichlet & 11 & 30s & 336s \\
  & & TV & 11 & 36s & 392s \\ 
   \hline
\end{tabular}
\end{center}
\caption{Computational complexity of the proposed approaches using SBL data, i.e., with $L=1$ (timings in seconds).
}
\label{table:time_L1}
\vspace{-6mm}
\end{table}

Table \ref{table:time_L1} illustrates the computational cost of our method with SBL data, for the different models considered. The time corresponds to that the execution of Algo. \ref{algo:algo1} and does not include the estimation of $\textbf{t}$ which has a fixed complexity (for fixed $(N,L,T)$), that is 106s here. It however includes the estimation of the reflectivity profile from $\boldsymbol{W}$, which takes 0.25s and is negligible compared to the other steps. Since Lap (resp. G-Dirichlet) has a complexity similar to that of TV (resp. W-Dirichlet), the corresponding results are omitted. While the convergence speed is not significantly affected by the SBR or number of photons, the iterations become slower as the photon counts increase, mostly due to the cost of the likelihood evaluation. All the methods present comparable computational costs, although W-Dirichlet (resp. TV) is generally the fastest (resp. slowest) method with SBL data.

Based on these first results, we only consider C-Dirichlet, Lap and TV, as the most promising models, for the analysis of SW-MSL data in the next section. The W-Dirichlet model is also considered to illustrate the impact of the lack of informative prior models on the inference process.

\subsection{Analysis of real SW-MSL data}
\label{res:RD_SW-MSL}
Here, we compare our approach to that developed in \cite{Comp_realData}, using the real SW-MSL data acquired in that work, composed of $L=4$ wavelengths ($473, 532, 589$ and $640$nm), whose IRFs are shown in Fig. \ref{Figurine} (see \cite{Comp_realData} for additional details about the imaging setup). While perfect ground truth reflectivity and range profiles are not available, we use as reference profiles the results obtained with the method developed in \cite{Comp_realData}, applied to data acquired with a long illumination time and a negligible background contribution. We consider datasets with 1.1, 5.7, 11.4 and 114.3 signal photons per pixel on average. Two scenarios are considered. One set of data was acquired with negligible background illumination while a second data set was acquired with spatially varying background (see \cite{Comp_realData}), leading to an average SBR of 1.4. 

\begin{figure}[ht!]
    \begin{center}
\resizebox{88mm}{!}{
    \includegraphics{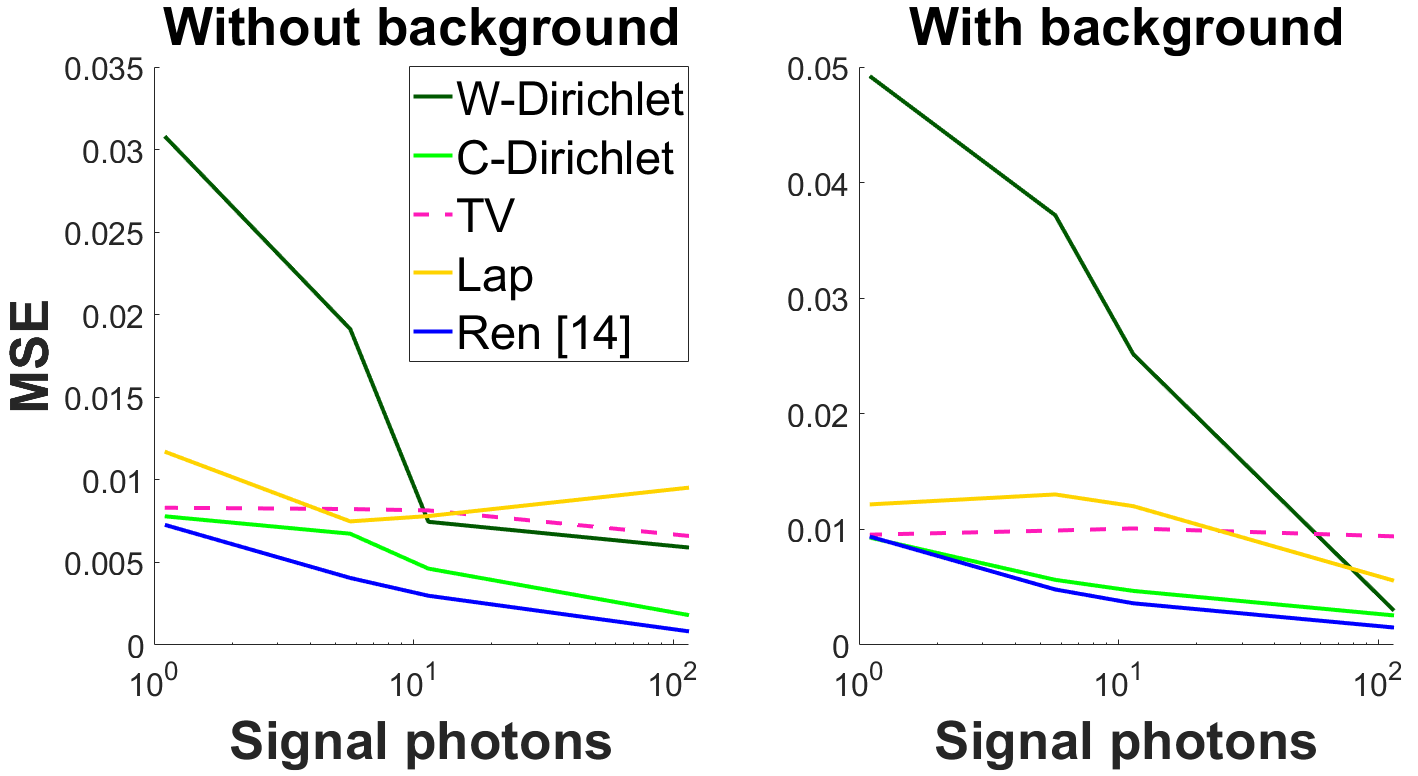}}
    \end{center}
    \caption{MSE of $\boldsymbol{R}$ obtained with real data ($L=4$) and the different competing methods as function of the number of photon counts, with (left) and without background illumination (right). \label{fig:MSE_RD} }
\end{figure}

The reflectivity MSEs obtained by the different competing methods are shown in Fig. \ref{fig:MSE_RD} for different acquisition scenarios, i.e., number of signal photons. 
Irrespective of the background level, C-Dirichlet is the method whose performance is the closest to the method from \cite{Comp_realData}. This figure also highlights the importance of prior information, especially in the presence of strong background. Note that TV and Lap perform generally worse than C-Dirichlet, especially when the photon counts are large. This is mainly due to the value of $\lambda$, which has been fixed for all the illumination conditions.
Although C-Dirichlet does not outperform the method of Ren \cite{Comp_realData}, the performance degradation is not significant and is balanced by the significantly lower computational cost, as will be shown at the end of this section.

\begin{figure}[!t]
\includegraphics[width=\columnwidth]{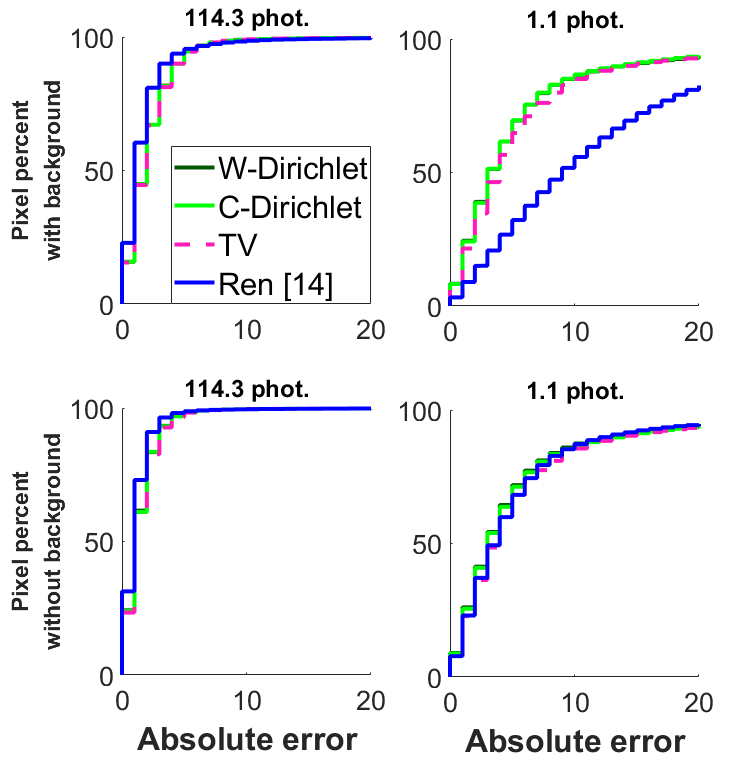}
\caption{CDFs of the depth absolute error, with high background ($\textrm{SBR}=1.4$) (top) and with negligible background (bottom). The first (resp. second) column corresponds to 114.3 (resp. 1.1) signal photons per pixel on average.}
\label{fig:Res_RD_CDF}
\end{figure}
\begin{figure}[!t]
  \includegraphics[width=\columnwidth]{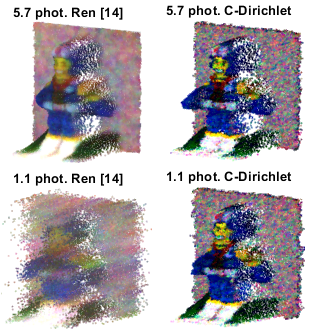}
    \caption{Depth/RGB reconstruction using real SW-MSL data, with the method of \cite{Comp_realData} and our approach (C-Dirichlet), with 1.1 (bottom) and 5.7 (top) signal photons per pixel, with non-negligible background (SBR$=1.4$). 
    \label{fig:Res_RD_3Dplot}}
\end{figure}


The depth error CDFs of the different methods, obtained with 114.3 and 1.1 signal photons per pixel, are shown in Fig. \ref{fig:Res_RD_CDF}. In this figure, Lap is omitted as it provides nearly the same results as TV. As in the SBL case (see Fig. \ref{fig:CDF_1D}), the ranging results obtained with our method do not strongly depend on the choice of the prior model for $\boldsymbol{W}$. 
With high photon counts (Fig. \ref{fig:Res_RD_CDF}, left column), the method of Ren \cite{Comp_realData} provides slightly better results, while in the low illumination regime, our method performs similarly or even better (see Fig. \ref{fig:Res_RD_CDF}, top-right subplot), due to the slow convergence of the method \cite{Comp_realData} in such cases. Conversely, our method which first marginalizes $\boldsymbol{t}$ converges quickly and is more robust in the photon-starved and high background regime.
For completeness, the estimated depth profiles obtained by Ren \cite{Comp_realData} and our approach using C-Dirichlet for $1.1$ and $11.4$ signal photons per pixel, and with strong ambient illumination are shown in Fig. \ref{fig:Res_RD_3Dplot}. This figure illustrates the overall agreement between the two methods, except in the most extreme scenario where our new method provides a more accurate range profile. It also shows that the C-Dirichlet leads less blurred reflectivity profiles.

\begin{table}[htb]
\begin{center}
 \begin{tabular}{|c|c|c|c|c|c|}
 \hline
  Counts /  & \multirow{2}{*}{Method} & \multirow{2}{*}{$N_{iter}$} & Time /& Total \\
  SBR & & & iter & time \\
 \hline
   \multirow{3}{*}{1.1/$\infty$} & W-Dirichlet & 11 & 35s & 386s \\
  & C-Dirichlet & 11 & 36s & 396s \\
  & TV & 11 & 413s & 4546s \\ 
   \hline
   \multirow{3}{*}{1.9/1.4} & W-Dirichlet & 12 & 34s & 433s \\
  & C-Dirichlet & 12 & 37s & 445s \\
  & TV & 12 & 381s & 4574s \\ 
   \hline
   \multirow{3}{*}{114.3/$\infty$} & W-Dirichlet & 10 & 53s & 531s \\
  & C-Dirichlet & 11 & 56s & 617s \\
  & TV & 11 & 581s & 6391s \\ 
   \hline
   \multirow{3}{*}{194/1.4} & W-Dirichlet & 13 & 55s & 716s \\
  & C-Dirichlet & 11 & 58s & 639s \\
  & TV & 11 & 609s & 6702s \\ 
   \hline
   Any & Ren & NA & NA & $\approx$ 15 hours \\
   \hline
\end{tabular}
\end{center}
\caption{Computational cost of the competing approaches for real SW-MSL data analysis, for different total counts and SBRs.}
\label{table:time_L4}
\vspace{-6mm}
\end{table}

Finally, Table \ref{table:time_L4} reports the computational time associated with the different prior models and the method from \cite{Comp_realData}. The results related to the Lap are similar to those of TV and are omitted here. Note that the table does not include the estimation of $\boldsymbol{t}$, which equals $112$s, irrespective of the background level and signal photon counts. It does however include the estimation of $\boldsymbol{R}$ from $\boldsymbol{W}$, which lasts 0.25s and does not depend on $L$.
While the two Dirichlet-based models have almost the same complexity, the TV-based model has a higher computational cost due to the use of an ADMM to solve \eqref{eq:modified_M}, whose complexity grows with $L$.

The results confirm the benefits of the C-Dirichlet model, which generally achieves similar or better depth estimation and satisfactory spectral signatures reconstruction, when compared to the existing method. This is however performed at a significantly lower computational cost.



\section{Conclusion} \label{Sec:Conclusion}
In this paper, we developed a novel algorithm for spectral and depth profile estimation from SW-MSL waveforms, in the presence of non negligible background and reduced acquisition time (down to 1 signal photon per pixel, on average). The reformulation of the model into a mixture of illumination sources and ambient illumination allowed the use of an EM-based algorithm for sequential parameter estimation, allowing in turn a significant speed-up compared to the existing method for SW-MSL data analysis, without significant degradation of the reconstruction performance. We compared several reflectivity models and demonstrated the benefits of the C-Dirichlet model in terms of quality of estimation, level of user supervision and computational cost.
While our new approach benefits from a significantly reduced computational cost, these new results are still incompatible with real-time requirements. Thus, future work should consist of further accelerating the inference process. An interesting route for improvement could be online reconstruction, as in \cite{altmann2019fast}. 


\section*{Acknowledgement}
We thank the single-photon group led by Prof. G. S. Buller for providing the real single-photon data used in this work. We gratefully acknowledge the support of NVIDIA Corporation with the donation of the Titan XP GPU used in this research.

\bibliographystyle{ieeetr.bst}
\bibliography{biblio}

\end{document}